\documentclass[preprint,12pt]{elsarticle}
\usepackage{amssymb}
\usepackage{centernot}
\usepackage{lipsum}
\usepackage[font=footnotesize,labelfont=bf]{caption}
\usepackage{subcaption}
\usepackage{float}
\usepackage{csquotes}
\usepackage{amsmath}
\usepackage{fancyhdr}
\usepackage{blindtext}
\DeclareMathOperator{\sech}{sech}

\journal{Nuclear Physics B}

\begin{document}
\begin{frontmatter}

\author{Irfan Mahmood\corref{cor1}\fnref{a,b}}
\ead{mahirfan@yhaoo.com}

\affiliation[a]{organization={Department of Mathematics, },
             addressline={ Colllege of Science, Shanghai University},
             city={Shanghai},
             country={200444, P.R.China }}

\author{,  Hira Sohail\corref{cor1}\fnref{b}}
\ead{hirasohail02@outlook.com}

\affiliation[b]{organization={Centre for High Energy Physics},
             addressline={University of the Punjab},
             city={Lahore},
             postcode={54590},
             state={Punjab},
             country={Pakistan}}
             
\author{and Allah Ditta \corref{cor1}\fnref{a}}
\ead{mradshahid01@gmail.com}

\title{Darboux Wronskian solutions of Ito typed coupled KdV equation   with exact solitonic solutions and conserved densities }

\begin{abstract}

In this article, we derive the Darboux solutions of Ito type coupled KdV equation in Darboux framework which is associated with Hirota Satsuma systems. Then we generalise  $N$-fold Darboux transformations in terms of Wronskians. We also derive the exact multi-solitonic solutions for the coupled field variables of that system  in the background of zero seed solutions. The last section encloses the derivation of continuity equation with several  conserved densities through its Riccati equation. 
\end{abstract}

\end{frontmatter}

\section{Introduction}
The nonlinear  solitonic equations have got considerable attentions in theory of integrable systems because of their wide applications in various domains of  physics and applied mathematics. One of the very interesting and earliest solitonic equations is KdV equation that plays a crucial  role in the study of hydrodynamics to describe the geometrical properties of wave propagations in shallow water \cite{1}  and has also been acknowledged  as integrable model in the analysis of electron plasma waves phenomenon associated with   cylindrical plasma system  \cite{2}. The study of  solitonic solutions of nonlinear evolution equations attained substantial importance  in modern theory of integrable systems in exploring  algebraic and  geometrical profiles with their physical aspects, for example in context of Bose-Einstein condensate different types of solitons such as bright solitons  \cite{3, 4} dark solitons  \cite{4}, vortex solitons  \cite{5} and gap solitons  \cite{6} have been found while studying  matter wave solitons.
In this article we construct the solitonic solutions  of  Ito type coupled KdV equation  wave equation
 \begin{equation}\label{1} 
\left\{
\begin{array}{lr}
   u_t=u_{xxx}+6uu_x+ q_x \\
     v_t=2(uv)_x
\end{array}
\right.
 \end{equation} 
which also called Integrable coupled nonlinear wave. The  above  coupled nonlinear wave  (CNW) equation  (\ref{1})  with $q=v^2 $ can be obtained  as the parametric reduction of  famous Ito system  \cite{7}  has been applied as integrable model in various domains of physics and fluid mechanics.  Recently,  its integro-differential  analogue \cite{8} with  mixed dark-bright solitonic solutions have been investigated.  Moreover, that coupled nonlinear wave equation has earned much importance in theory of integrable systems as it involves the KdV structure and  reduces to the ordinary KdV equation by setting the coupling variables $v$  as $v=0$. Here, we apply  Darboux transformation  to investigate its solitonic solutions. That Darboux  approach   \cite{matveev1991darboux, gu2004darboux, trisetyarso2009application} has been acknowledged as one of the efficient tools in theory of integrable systems to calculate the exact solutions of these systems with their  algebraic and geometrical properties.  In literature number of successful implementations of this transformation have been shown as an efficient   integrable tool   which enhance its significance from physical point of views. Among these number of remarkable applications of DT few of them are mentioned here as applied in the analysis of electrodynamical features \cite{TA2010} in case of  quantum cavity problems and also to investigate the geometrical properties of graphene  \cite{trisetyarso2011erratum} with exact solitonic solutions. Moreover, that method has also been applied fruitfully to construct the quasideterminant solutions of the Painlevé II equation \cite{irfan2012lax} with its related Toda system \cite{mahmood2015quasideterminant} for its non-commutative analogs and to derive  the exact solutions of the generalized coupled dispersionless integrable system\cite{hassan2009darboux}. 
In section $2$, we construct  one-fold,  two-fold and three-fold Darboux solutions of  equation  (\ref{1}) with the help of linear representation for its  coupling field variable $u$ and $v$ and then we generalise  its  $N$-fold Darboux solutions in determinantal form as in terms of  Wronskians.

Subsequently, we derive  exact multi-solitonic solutions  in the background of zero seed solutions for both coupling field variables $u$ and $q$ with graphical presentations of variety of solutions  scattering elastically. This work also encloses the derivation of matrix zero curvature representation of  CNW equation (\ref{1}) possessing traceless matrices through its existed  scalar Lax pair. That representation may be assumed to fit in the AKNS scheme   \cite{AKNS} as it usually involves the parametric  traceless matrices of  order $N$ containing field variables. In last section we derive equation of continuity through the linear representation of   CNW equation (\ref{1}).  We also  calculate its several  conserved densities with the help of its Riccati equation. 

\section{Linear representations  and  The Darboux solutions}
This section encloses the Lax representation of Ito type coupled KdV equation which also called Integrable coupled nonlinear wave (CNW) equation  (\ref{1}) and the derivation of its equivalent zero-curvature from scalar Lax pair. Then by using the  Darbaoux transformation \cite{matveev1991darboux, gu2004darboux} on arbitrary function we  construct the one-fold, two-fold and three-fold Darboux solutions to coupling field variables $u$ and $v$ in terms of seed solutions. Subsequently we generalize their $N$-fold Darboux solutions in determinantal form.
\subsection{Lax Pair and Zero-curvature representation}
The coupled nonlinear wave equation (\ref{1}) possesses is integrable  \cite{AV, kai2021exact}  and arises as the compatibility of  subsequent linear system   

    \begin{equation}          \psi_{xx}=\left(\lambda-u-\beta q\right) \psi.\\
        \label{2}
    \end{equation}
        
    \begin{equation} 
         \psi_t=\left(4 \lambda+2u\right)\psi_x -u_x \psi.
       \label{3}
    \end{equation}
where $q=v^2$ , $\beta = \frac{1}{4 \lambda}$ and  $\lambda$ is spectral parameter. It is easy to obtain CNW equation  (\ref{1}) by elimination of arbitrary function $ \psi_{xx}$ from above linear system.
\subsection*{Proposition 2.1}
By introducing a  column vector in terms of arbitrary function as $ F=(f_0,f_1)^T=(\psi,\psi_x)^T$, we may construct the matrix zero-curvature representation of CNW equation  (\ref{1}).
\subsection*{Proof}
Consider the following linear system  for arbitrary function $F$
\begin{equation}\label{4} 
\left\{
\begin{array}{lr}
  \partial_xF=AF \\
 \partial_tF=BF
\end{array}
\right.,
 \end{equation}
 where $A$ and $B$ are the matrices of order $2$ to be determined, the compatibility of  linear system  (\ref{4})  yields zero-curvature form as below
 \begin{equation}\label{5}
  \partial_t A-\partial_x B+[A,B]=0,
 \end{equation}
This can be shown that with the help of $ F=(f_0,f_1)^T=(\psi,\psi_x)^T$, the eigenvalue equation  $ L\psi=\lambda \psi$
will take the following form
 \begin{align}\label{6}
    \partial_x (f_0,f_1)^T
    =A
    (f_0,f_1)^T,
\end{align}
where $ A=\begin{bmatrix}
       0&1\\
       \lambda-u-\beta q&0
    \end{bmatrix}$ and can also be written as $F_x=AF$.
For the temporal part, let take the derivatiion  of $ \psi_t=\left(4 \lambda+2u\right)\psi_x -u_x \psi$  with respect to $x$, then by using $ \psi_{xx}=\left(\lambda-u-\beta q\right) \psi$ in resulting expression , we get
\begin{equation}\label{7}    \psi_{tx}=u_x\psi_x+4\lambda^2\psi-2\lambda u\psi-v^2\psi-2u^2\psi-\frac{uv^2 \psi}{2\lambda}-u_{xx}\psi,
\end{equation}
Now combining equation  (\ref{3}) and above equation, we obtain  $ F_t=BF$, where 

\begin{equation}\label{8}
    B=\begin{bmatrix}
    -u_x&4\lambda+2u\\
    4\lambda^2-2\lambda u-v^2-2u^2-\frac{uv^2}{2\lambda}-u_{xx}&u_x
    \end{bmatrix},
\end{equation}
\subsection*{Remark}
The above matrix zero-curvature representation is derived from its scalar analogue presented in  \cite{AV} whose compatibility condition is equivalent to ito type system as  CNW equation (\ref{1})   and the method mentioned above is straight forward to calculate the matrix zero curvatue representation  from existed scalar Lax pair. 
\subsection{Darboux solutions}
Here we construct the explicit Darboux expressions for the field variables $u$ and $v$ which connect the old solutions of CNW equation (\ref{1}) to its new solutions through the particular solutions of linear equations (\ref{2}) and (\ref{3}). In order to construct the Darboux solutions for the coupled field variables $u$ and $v$, let $u[1]$ and $q[1]$ are new solutions of CNW equation (\ref{1}) and  $\psi[1]$ is also a new solution of its associated linear system, then linear equation (\ref{2})  with these new solution becomes
  \begin{equation} 
    \psi_{xx}[1]=\left(\lambda-u[1]-\beta q[1]\right)\psi[1],
    \label{9}
\end{equation}
the transformation on arbitrary function is defined as
\begin{equation} 
    \psi[1]=\psi_x - \sigma_{1} \psi,
    \label{10}
    \end{equation}
 where $\sigma_{1}=\frac{\psi_{1x}}{\psi_1}$  and the $\psi_1$ is a particular solution can be calculated at $\lambda=\lambda_{1x}$ from linear system (\ref{2}) and (\ref{3}) with provided seed solutions.
Now substitute the value of $\psi[1]$ from equation (\ref{10}) into transformed expression (\ref{9}) then by using the original linear system (\ref{2}), we can  extract one step Darboux transformation on $u$ and $q$ as below

\begin{equation} 
 u[1]=u+2\sigma_{1x},
 \label{11}
\end{equation}
\begin{equation}\label{12}
    q[1]=q-q_x \sigma_1^{-1},
\end{equation}
respectively. In above transformations  $ u[1]$ and $ q[1]$ are the new solutions generated from old solutions $u$ and $q$ through the particular solutions of linear system (\ref{2}) and (\ref{3}). This can be shown that with trivial solutions $u=0$, $q=0$, there is no variation on $q[1]$ that remains trivial and this problem can be eliminated by substituting the value of $q_x$ from first equation of system (\ref{1}), then finally we  can express transformation (\ref{12}) in following form
\begin{equation}\label{13}
    q[1]=q- KdV(u) \sigma_1^{-1},
\end{equation}
where $KdV(u)=u_t-u_{xxx}-6uu_x$ which generates non-trivial solutions for $ q[2]$, $ q[3]$, ..., and so on.
\subsection*{Remark}
For particular case, taking coupling field variable $v$ as constant the Darboux transformation (\ref{11}) of KdV equation for fixed seed solution yields  its real valued rational solutions  as discussed in \cite{VB2} for real  parametric particular solutions and this  transformation can also be applied to generate its special class of solutions as positon solutions for periodic  generating function $\sigma$. Here our focus is to  discuss only about multi-solitonic solutions CNW (\ref{1}) with their conserved densities therefore we omit here to incorporate the non-solitonic behaviours.
\subsection{Two and Three Fold Darboux transformations}

The two-fold Darboux transformation for  arbitrary fuction  $\psi$ can be written as 

\begin{equation}
    \psi[2]=\left(\frac{d}{dx}-\frac{\psi_{2x}[1]}{\psi_2[1]}\right)\left(\frac{d}{dx}-\frac{\psi_{1x}}{\psi_1}\right)\psi, 
    \label{15} 
\end{equation}
and   $\psi_2[1]$  can be calculated from following expression 
\begin{equation}
    \psi_2[1]=\psi_{2x}-\frac{\psi_{1x}}{\psi_1}\psi_2,
\end{equation}\label{n-1}
where $\psi_2$ is the particular solution  linear systems (\ref{2})  and (\ref{3}) at  $\lambda = \lambda_2$, simply we can write two-fold transformation (\ref{15}) as ratio of Wronskians 
\begin{equation}
    \psi[2]=\frac{W(\psi_1,\psi_2,\psi)}{W(\psi_1,\psi_2)}.
    \label{16}
\end{equation}
\begin{equation}
W(\psi_1,\psi_2,  \psi)=
\begin{vmatrix}
  \psi_1&\psi_2  &\psi \\
  \psi_1^{(1)}&\psi_2^{(1)} &\psi^{(1)}\\
   \psi_1^{(2)}&\psi_2^{(2)}&\psi^{(2)}\\
 \end{vmatrix}, W(\psi_1,\psi_2 )=
\begin{vmatrix}
  \psi_1&\psi_2  \\
  \psi_1^{(1)}&\psi_2^{(1)}\\
   \end{vmatrix}
\end{equation}

here the superscripts $(i)$ represent the order of derivatives. Similarly, we can construct the  $  \psi[3]=\left(\frac{d}{dx}-\frac{\psi_{3x}[2]}{\psi_3[2]}\right)\psi[2]$  in terms of Wronskians as below 
\begin{equation}    \psi[3]=\frac{W(\psi_1,\psi_2,\psi_3,\psi)}{W(\psi_1,\psi_2,\psi_3)},
    \label{n-2}
\end{equation}
where 
\begin{equation}
W(\psi_1,\psi_2,\psi_3,  \psi)=
\begin{vmatrix}
  \psi_1&\psi_2 & \psi_3  &\psi \\
  \psi_1^{(1)}&\psi_2^{(1)}& \psi_3^{(1)}&\psi^{(1)}\\
   \psi_1^{(2)}&\psi_2^{(2)}& \psi_3^{(2)}&\psi^{(2)}\\
  \psi_1^{(3)}&\psi_2^{(3)}& \psi_3^{(3)}&\psi^{(3)}\\
\end{vmatrix}
\end{equation}
\begin{equation}
W(\psi_1,\psi_2,\psi_3 )= \begin{vmatrix}
  \psi_1&\psi_2 & \psi_3  \\
  \psi_1^{(1)}&\psi_2^{(1)}& \psi_3^{(1)}\\
   \psi_1^{(2)}&\psi_2^{(2)}& \psi_3^{(2)}\\
  \psi_1^{(3)}&\psi_2^{(3)}& \psi_3^{(3)}\\
\end{vmatrix}.
\end{equation}
Here we have  presented Darboux transformations for arbitrary function $\psi$ upto three-fold  in terms of Wronskians. In following proposition 2.2, we elaborate a procedure \cite{matveev1991darboux, DCT}  to generalise the $N$-fold transformation for the coupling  field variables $u$ and $q$ in compact form as the logarithmic derivative of  Wronskians.   
\subsection*{Proposition 2.2}
With the help of one-fold Darboux transformations, we can construct $N$-fold Darboux transformations  for coupling field variables  in following compact forms     

\begin{equation}
    u[N]=u-2\eta_{1x}, 
    \label{22} 
\end{equation}
and 
\begin{equation}\label{12}
    q[N]=q-q_x \eta_{1}^{-1},
\end{equation}
here $ \eta_{1}= \frac{d}{dx}\log{W\left(\psi_1,\psi_2,...,\psi_N\right)}$ and $W(\psi_1,\psi_2,......,\psi_N)$ is  Wronskian of order $N$.\\
\textbf{Proof: }\\
The second iteration of (\ref{10}) yields  two-fold Darboux transformation  on $ \psi $ 
\begin{equation}
    \psi[2]=\left(\frac{d}{dx}-\sigma_2\right)\psi[1]=\left(\frac{d}{dx}-\sigma_2\right)\left(\frac{d}{dx}-\sigma_1\right)\psi
\end{equation}
with $ \sigma_2=\psi_2^{\prime}[1]\psi_2^{-1}[1] $ and $\psi_2[1]$ can be calculated as $   \psi_2[1]=\left(\frac{d}{dx}-\sigma_1\right)\psi_2$, here  $\psi_2$ is the particular solution of linear systems (\ref{2}) and (\ref{3}) at $\lambda=\lambda_2$. Similarly the two-fold Darboux transformation on $ u$  and $q$ respectively can be written as
\begin{equation}\label{2FD}
    u[2]=u+2\sigma_{1x}+2\sigma_{2x}
\end{equation}
\begin{equation}\label{2FD2}
    q[2]=q- KdV(u) \sigma_1^{-1}- KdV(u[1]) \sigma_2^{-1}.
\end{equation}
After $N$ iteration we obtain the $N$-fold Darboux transformations as below
\begin{equation}\label{NFD}
    \psi[N]=\left(\frac{d}{dx}-\sigma_N\right)\left(\frac{d}{dx}-\sigma_{N-1}\right)\left(\frac{d}{dx}-\sigma_{N-2}\right)...\left(\frac{d}{dx}-\sigma_2\right)\left(\frac{d}{dx}-\sigma_1\right)\psi
\end{equation}
\begin{equation} \label{NFD2}
    u[N]=u+2  \sum_{i=1}^{N}  \sigma^{\prime}_{i}.
\end{equation}
\begin{equation}\label{NFD3}
  q[N]=q-  \sum_{i=1}^{N} KdV(u[i-1]) \sigma_i^{-1}.
\end{equation}
 with $ \sigma_i=\psi^{\prime}_i[i-1]\psi^{-1}_i[i-1]$. 
 Now in order construct  $N$-fold Darboux transformations in terms of Wronskian, we start with $N$-fold Darboux transformation for arbitrary function $\psi$ in following form
\begin{equation}\label{NFS}
    \psi[N]=D[N]\psi=\psi^{(N)}+\eta_1 \psi^{(N-1)}+\eta_2 \psi^{(N-2)}+.....+\eta_{N-1} \psi^{(1)}+\eta_N \psi.
\end{equation}
which is an equivalent representation of   (\ref{NFD}), now we can  easily show that the  linear system  (\ref{11})  under the $N$-fold transformation  $(\ref{NFS}) $ may lead the following transformations 
\begin{equation}\label{NFDU}
    u[N]=u-\eta_{1x},  q[N]=q-q_x \eta_{1}^{-1}.
\end{equation}
The coefficient $\eta_1$ in  $N$-th order linear differential operator $D[N]$ can be determine from the following system of  $N$ linear algebraic equations 
\begin{equation} \label{LSA}
     \sum_{i=1}^{N} \eta_i \psi^{N-i} = -\psi^{N}_i 
\end{equation}
where $i=1,2,3,......,N$ that  
which  can be derived by taking  $\psi = \psi _i $ as $i$-th particular  solution into the equivalent form of  $N$-fold Darboux representation of  (\ref{NFS}),  we can calculate the $ \eta_1$ in following form by applying the Kramer rule from above system  (\ref{LSA})
\begin{equation}
    \eta_1=-\frac{\begin{vmatrix}
        \psi_1&\psi^{\prime}_1&\cdots&\psi^{N-1}_1&\psi^N_N\\
        \vdots&\vdots&\ddots&\vdots&\vdots\\
        \psi_n&\psi^{\prime}_N&\cdots&\psi^{N-1}_N&\psi^N_N
    \end{vmatrix}}{W\left(\psi_1,\psi_2,.....,\psi_N\right)}=\frac{d}{dx}\log{W\left(\psi_1,\psi_2,...,\psi_N\right)}.
\end{equation}
Now we can express $N$-fold transformation (\ref{NFDU}) in terms of Wronskian as below  
\begin{equation}\label{NFDU1}
    u[N]=u-\frac{d^2}{dx^2}\log{W\left(\psi_1,\psi_2,...,\psi_N\right)}. 
\end{equation}
Now with the help of two-fold and three-fold transformations $\psi[2]$, $\psi[3]$,  we can directly generalise the  
$N$-fold expression for arbitrary function $\psi$ in terms of ratio of Wronskians as below
\begin{equation}
\psi[N]=\frac{W(\psi_1,\psi_2,......,\psi_N,\psi)}{W(\psi_1,\psi_2,......,\psi_N)},
    \label{21}
\end{equation}
here
\begin{equation}
W(\psi_1,\psi_2,....\psi_N)=
\begin{vmatrix}
  \psi_1&\psi_2 & \cdots &\psi_N \\
  \psi_1^{(1)}&\psi_2^{(1)}& \cdots  &\psi_N^{(1)}\\
  \vdots &\vdots & \ddots &\vdots\\
  \psi_1^{(n-1)}&\psi_2^{(n-1)}&\cdots& \psi_N^{(n-1)}
\end{vmatrix},
\label{24}
\end{equation}
and 
\begin{equation}
W(\psi_1,\psi_2,....\psi_N,\psi )=
\begin{vmatrix}
  \psi_1&\psi_2 & \cdots &\psi_N &\psi\\
  \psi_1^{(1)}&\psi_2^{(1)}& \cdots  &\psi_N^{(1)}&\psi^{(1)}\\
  \vdots &\vdots & \ddots &\vdots & \vdots  \\
  \psi_1^{(n-1)}&\psi_2^{(n-1)}&\cdots& \psi_N^{(n-1)}&\psi^{(n-1)}
\end{vmatrix},
\label{25}
\end{equation}
 in above determinants $\psi_j^{(i)}$ stands for $i$-th derivative of $\psi_j $ with respect to $x$ as $\psi_j^{(i)}= \frac{d^i \psi_j}{dx^i}$. 
In subsequent section,  we derive up to three soliton solutions for the coupling filed variables $u$ and $v$ in the background of zero seed solutions with their graphical representations.\\
\section{Exact solitonic solutions}
This section encloses the derivation of exact solutions to the CNW equation (\ref{1}) as one -soliton, two-soliton and three-soliton solutions for field variables $u$ and $v$ through  help of their  Darboux transformations in background of zero seed solutions.  In order to construct non-trivial exact solitonic solutions for field variable $v$, we apply the $N$ time iterative  form (\ref{NFD3}) for $q$ embedded with KdV equation.
\subsection{One-soliton solutions}
Let us start with simplest trivial solutions of CNW (\ref{1}) as $u=0$ and $v=0$ also $q=0$, then one-fold Darboux for $u$  will take the following form
\begin{equation} 
   u[1]=2\frac{d}{dx}\frac{\psi_{1x}}{\psi_1}=2\frac{d^2}{dx^2}\log\psi_1,
   \label{26}
\end{equation}
the particular solution $\psi_1$ can be calculated  from linear system (\ref{2}) at $\lambda =\lambda_{1}$ as below
\begin{equation}
    \psi_1(x,t)=2\cosh{(k_1x+4k_1 \lambda_1 t)},
    \label{27}
\end{equation}
Now after substituting the above value in expression (\ref{26}) and after some simplification we get one-soliton solution

\begin{equation}
     u[1]=2k_1^2 \sech^2{(k_1x+4k_1 \lambda_1 t)},
     \label{28}
\end{equation}
The dynamics of one-soliton solution in one dimension as well as on plane have been shown  respectively in following diagrams
\begin{figure}[H]
\begin{subfigure}{0.4\textwidth}
    \includegraphics[width=5cm ,height=5cm]{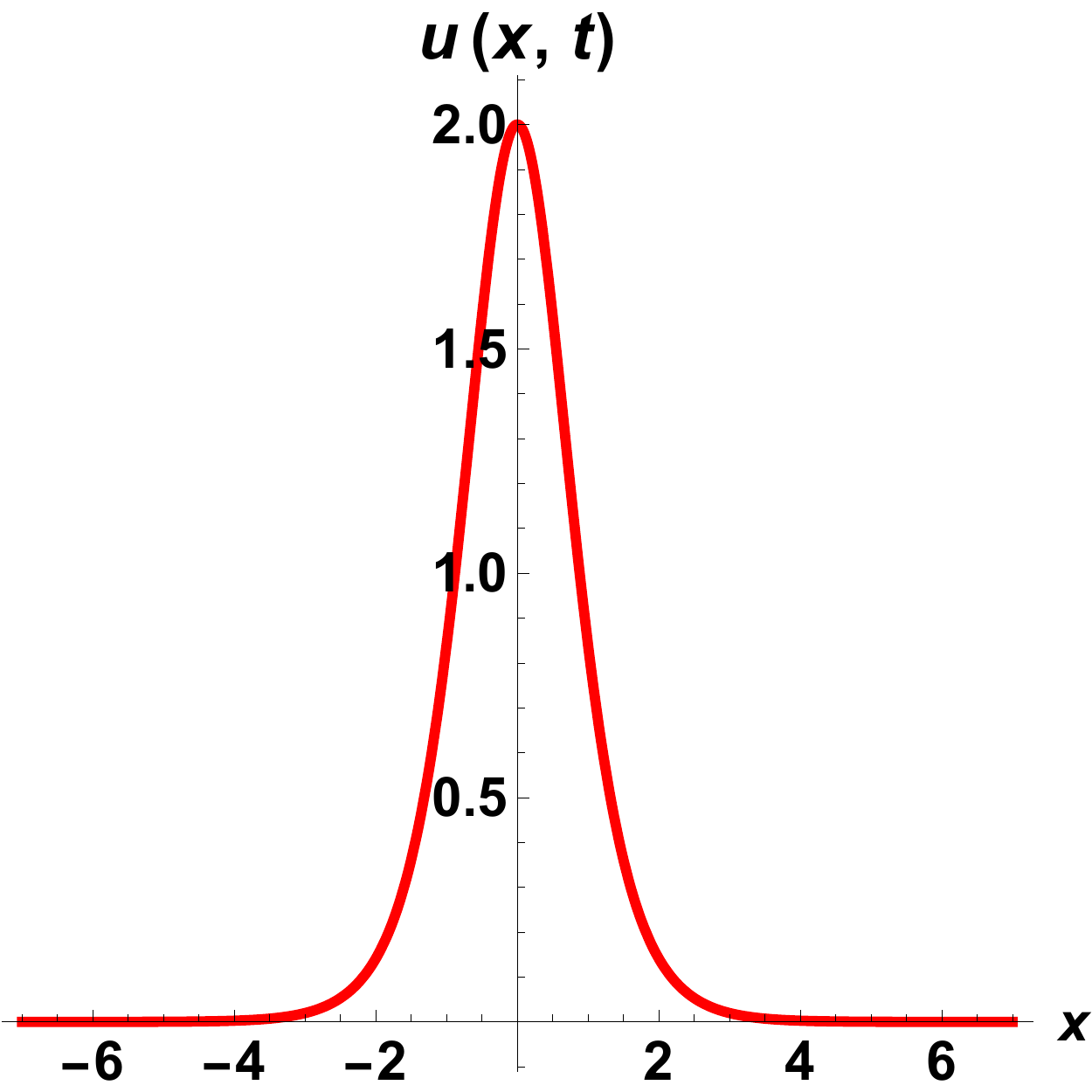}
    \caption{1-D}
    \label{Fig.1a}
\end{subfigure}
\hspace{1cm}
    \begin{subfigure}{0.4\textwidth}
    \includegraphics[width=5cm, height=5cm]{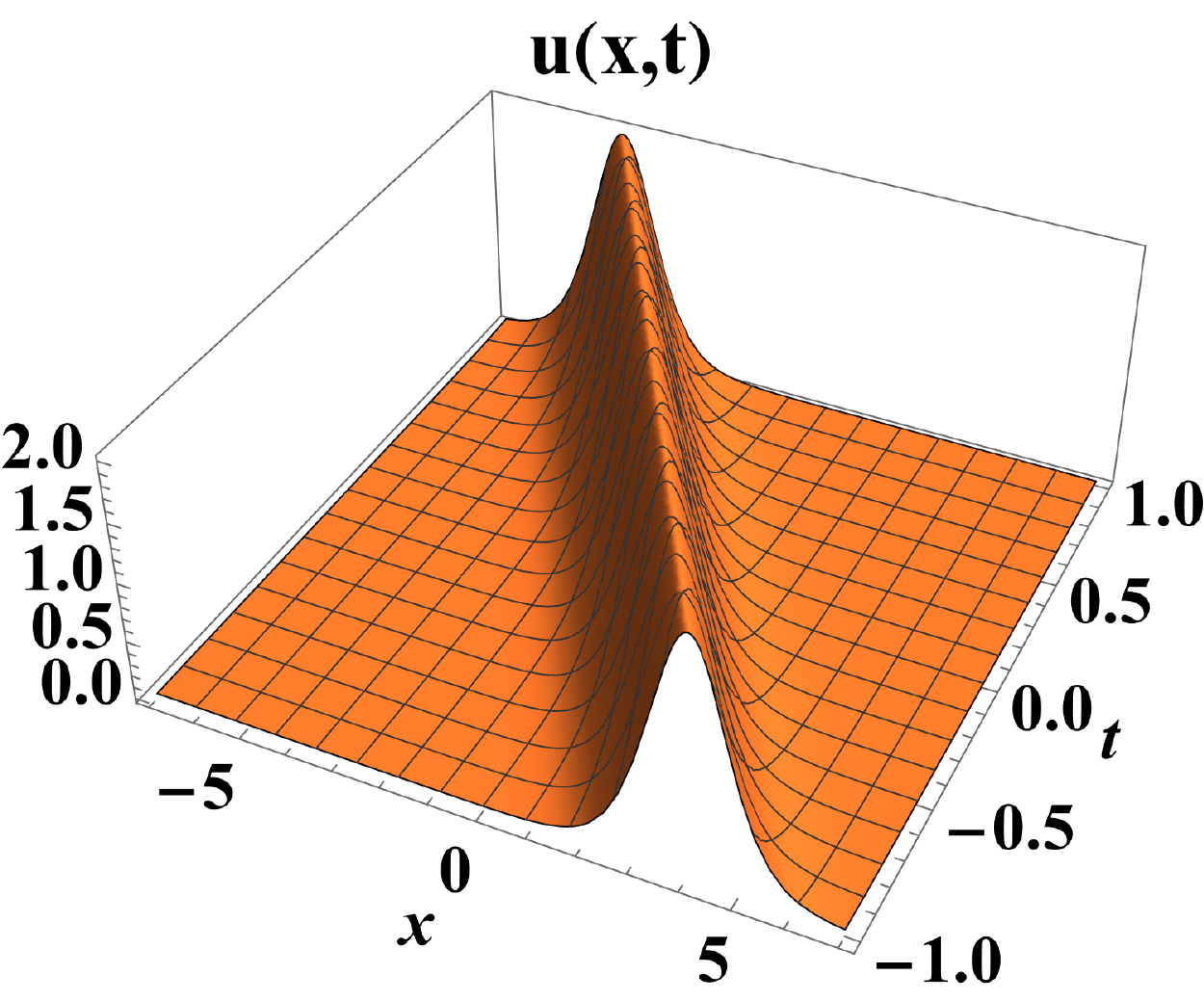}
    \caption{3-D}
    \label{Fig.1b}
    \end{subfigure}
    \caption{(a) represents the one dimensional dynamics of one soliton solution where as in (b)  its three dimensional profile has been shown.}
\label{Fig.1}
\end{figure}

\subsection{Two-soliton solution $u[2]$}
Now the two-fold Darboux transformation with trivial solution u=0 will take the following form 
\begin{equation}
    u[2]=2\frac{d^2}{dx^2}\log{[W(\psi_1,\psi_2)}],
   \label{29}
\end{equation}
where
\begin{equation}
    u[2]=2\frac{d^2}{dx^2}\log
    \begin{vmatrix}
    \psi_1&\psi_2\\
    \frac{d\psi_1}{dx}&\frac{d\psi_2}{dx}
    \end{vmatrix},
    \label{30}
\end{equation}
We can calculate the  value for $\psi_1$ and $\psi_2$ from the linear system (\ref{2}) at $\lambda=\lambda_1$ and $\lambda=\lambda_2$ respectively as below
\begin{equation}
    \psi_1(x,t)=2\cosh{(k_1x+4k_1 \lambda_1 t)},
    \label{31}
\end{equation}
\begin{equation}
    \psi_2(x,t)=2\sinh{(k_2 x+ 4 k_2 \lambda_2 t)},
    \label{32}
\end{equation}
After substituting above values into equation (\ref{30}), we get following results 

\begin{equation}
    u[2]=2\frac{d^2}{dx^2}\log
    \begin{vmatrix}
    2\cosh(k_1x+4 k_1\lambda_1 t)&2\sinh(k_2x+4 k_2\lambda_2 t)\\
    2k_1\sinh(k_1x+4 k_1\lambda_1 t)&2k_2\cosh(k_2 x +4 k_2\lambda_2 t)
    \end{vmatrix},
    \label{33}
\end{equation}
now expanding the determinate and taking derivations,  finally we obtain the following result  
\begin{equation}
    u[2]=4(k_2^2 - k_1^2)\left[\frac{k_2^2 \cosh(2\gamma_1)+k_1^2 \cosh(2\gamma_2)+k_2^2 - k_1^2}{((k_2-k_1)\cosh{(\gamma_1+\gamma_2)}+(k_2+k_1)\cosh{(\gamma_1-\gamma_2)})^2}\right],
    \label{34}
\end{equation}
which is two-soliton solution well know for KdV equation because KdV equation is embedded as essential part in CNW equation (\ref{1})  where $\gamma_1=k_1x+4 k_1\lambda_1 t$ and $  \gamma_2=k_2x+4 k_2\lambda_2 t$.
The inelastic scattering of the two solitons has been visualised  in one dimension as well as on plane as below
\begin{figure}[H]
\begin{subfigure}{0.4\textwidth}
    \includegraphics[width=1.2\linewidth, height=6cm]{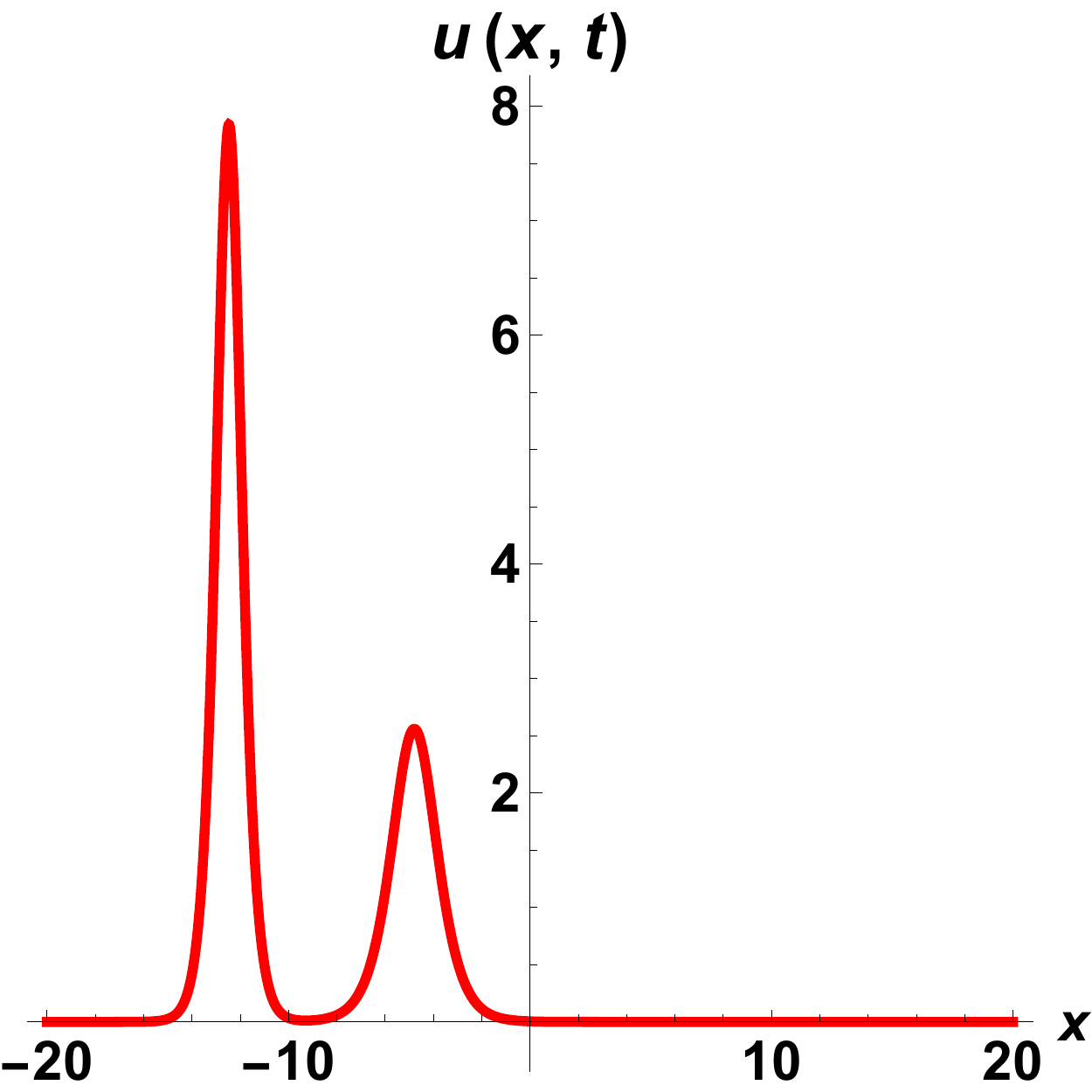}
    \caption{1-D}
    \label{Fig.3a}
\end{subfigure}
  \hspace{1cm}
    \begin{subfigure}{0.4\textwidth}
    \includegraphics[width=1.2\linewidth, height=6cm]{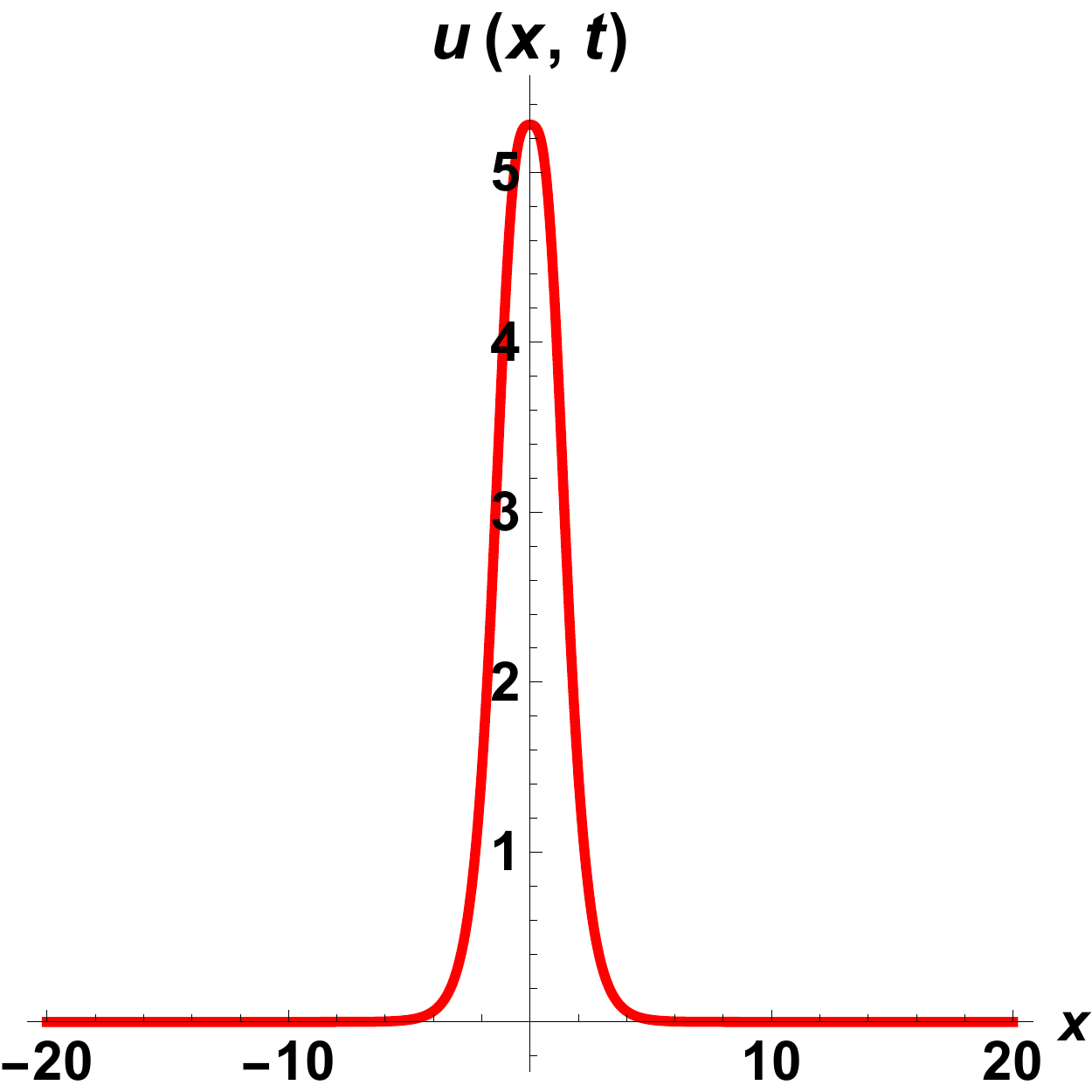}
    \caption{1-D}
    \label{Fig.3b}
    \end{subfigure}
    \caption{(a) represents the one dimensional dynamics of two soliton solution  before interaction where as in (b) one dimensional dynamics at the time of interaction has been shown.}
\label{Fig.3}
\end{figure}

\begin{figure}[H]
\begin{subfigure}{0.4\textwidth}
    \includegraphics[width=1.2\linewidth, height=6cm]{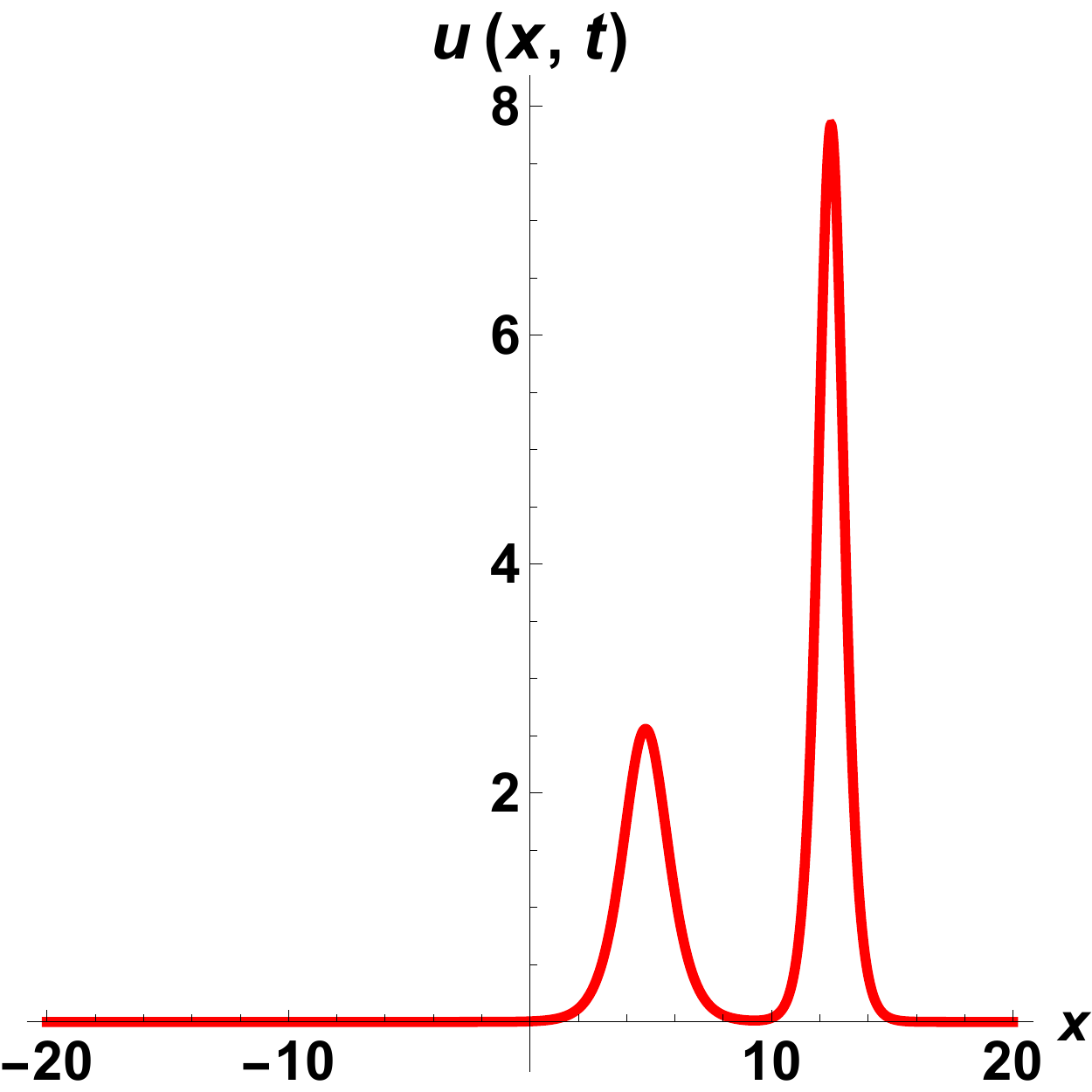}
    \caption{1-D}
    \label{Fig.4a}
\end{subfigure}
  \hspace{1cm}
    \begin{subfigure}{0.4\textwidth}
    \includegraphics[width=1.2\linewidth, height=6cm]{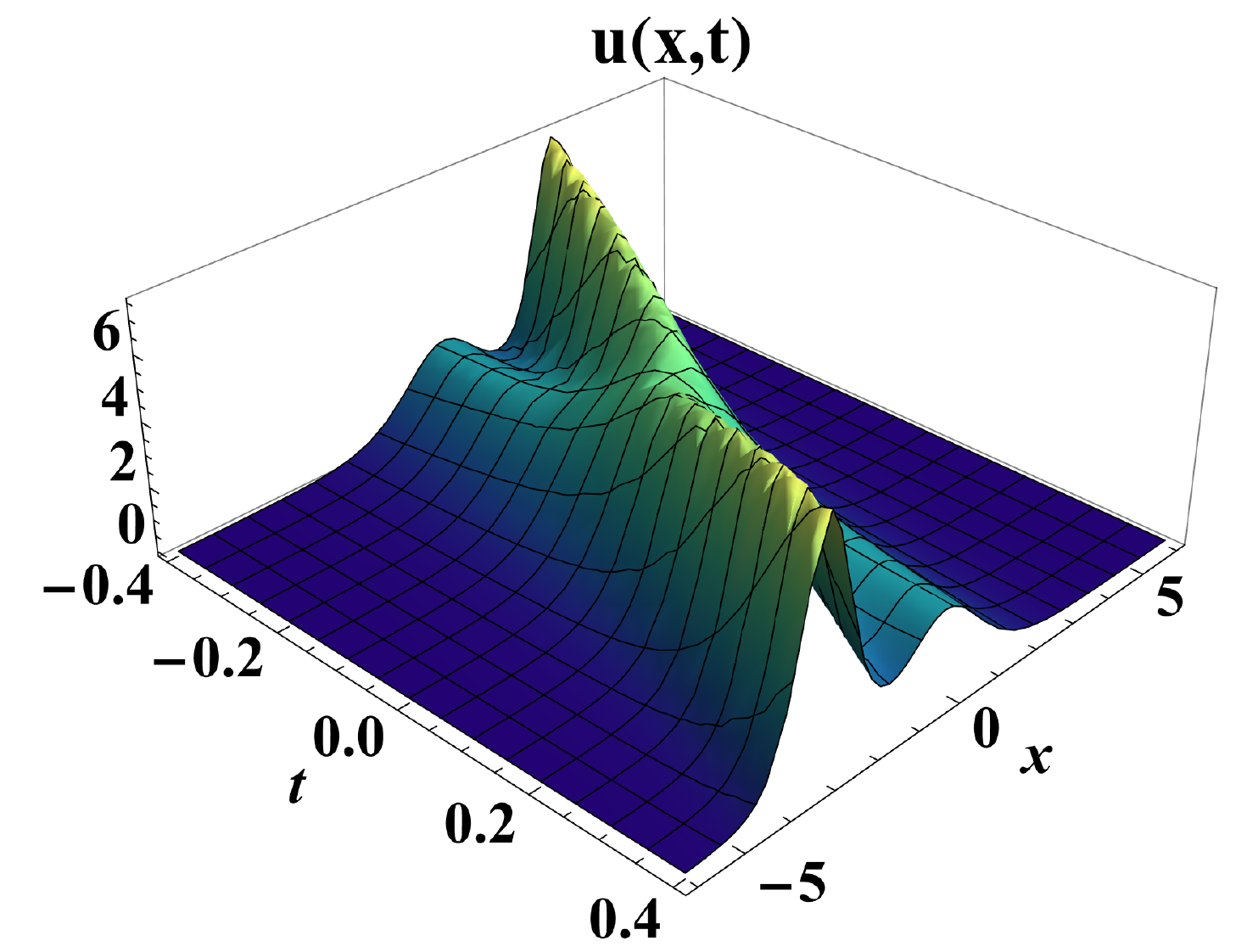}
    \caption{1-D}
    \label{Fig.4b}
    \end{subfigure}
    \caption{(a) represents the one dimensional dynamics of two soliton solution  after the interaction where as in (b) three dimensional dynamics of interaction has been shown.}
\label{Fig.4}
\end{figure}

\subsection{Two-soliton solution $q[2]$ for coupling variable $v$ }
In this part of section $2$, we derive two-soliton solution to coupling field variable $v$, as it is obvious from  $N$-the expression (\ref{23}) at $N=1$, $ q[1]=0$ implies $v[1]=0$, it means at first iteration only KdV solution appears. But the second iteration at $N=2$ yields non zero two soliton solution to coupling field variable $v$ as below

\begin{equation}\label{35}
    q[2]=q[1]-\left(u[1]_t-u[1]_{xxx}-6u[1]u[1]_x\right)\sigma_{2}[1]^{-1},
\end{equation}
Now substitute the calculated values of $u[1]$ and $\sigma_{2}[1]$  in last express (\ref{35}) and then after simplification, we get 
\begin{equation}\label{36}
 q[2]=\frac{ \left[16k_1^3 (k_1^2 - \lambda) \sech{\gamma_1}^2 \tanh{\gamma_1} (k_2 \coth{\gamma_2} - 
   k_1 \tanh{\gamma_1})\right]}{(k_1^2 - k_2^2)},
\end{equation}
this straight forward to obtain $ v[2]$ by taking square root of $ q[2]$ which is again two-soliton solution. The elastic interaction of these two solitons has been show in one dimension.
\begin{figure}[H]
\begin{subfigure}{0.4\textwidth}
    \includegraphics[width=5cm, height=5.5cm]{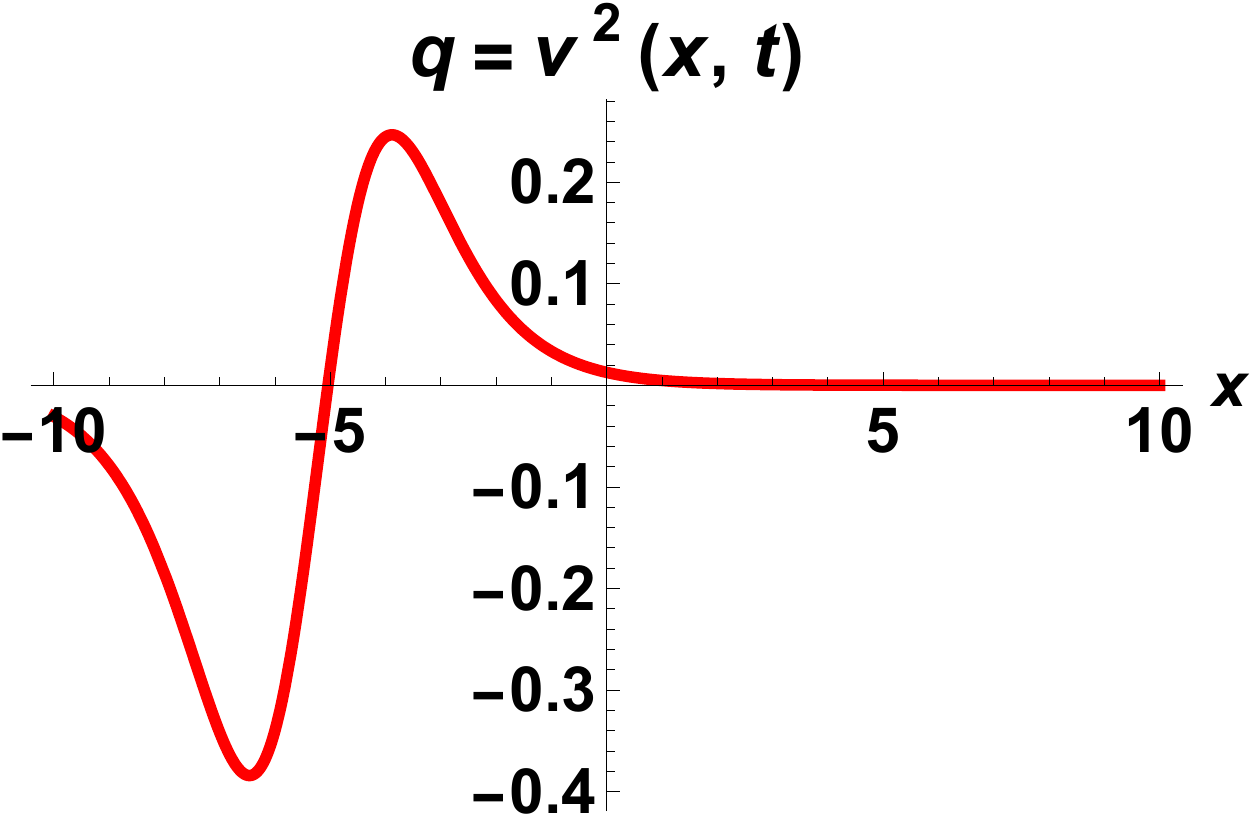}
    \caption{1-D}
    \label{Fig.6a}
\end{subfigure}
  \hspace{1cm}
    \begin{subfigure}{0.4\textwidth}
    \includegraphics[width=1.2\linewidth, height=6cm]{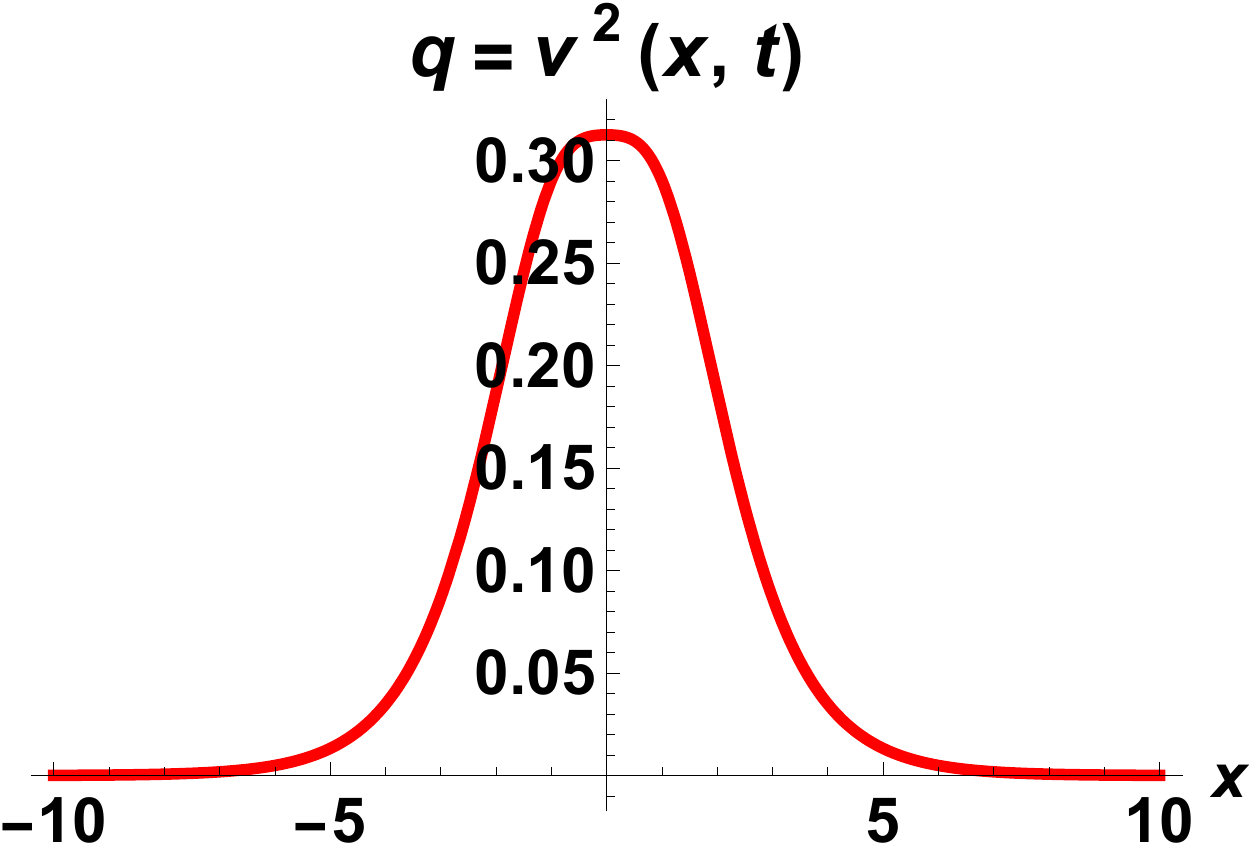}
    \caption{1-D}
    \label{Fig.6b}
    \end{subfigure}
    \caption{(a) represents the one dimensional dynamics of two soliton solution  before interaction where as in (b) one dimensional dynamics at the time of interaction has been shown.}
\label{Fig.6}
\end{figure}

\begin{figure}[H]
    \centering    \includegraphics[width=6cm, height=6cm]{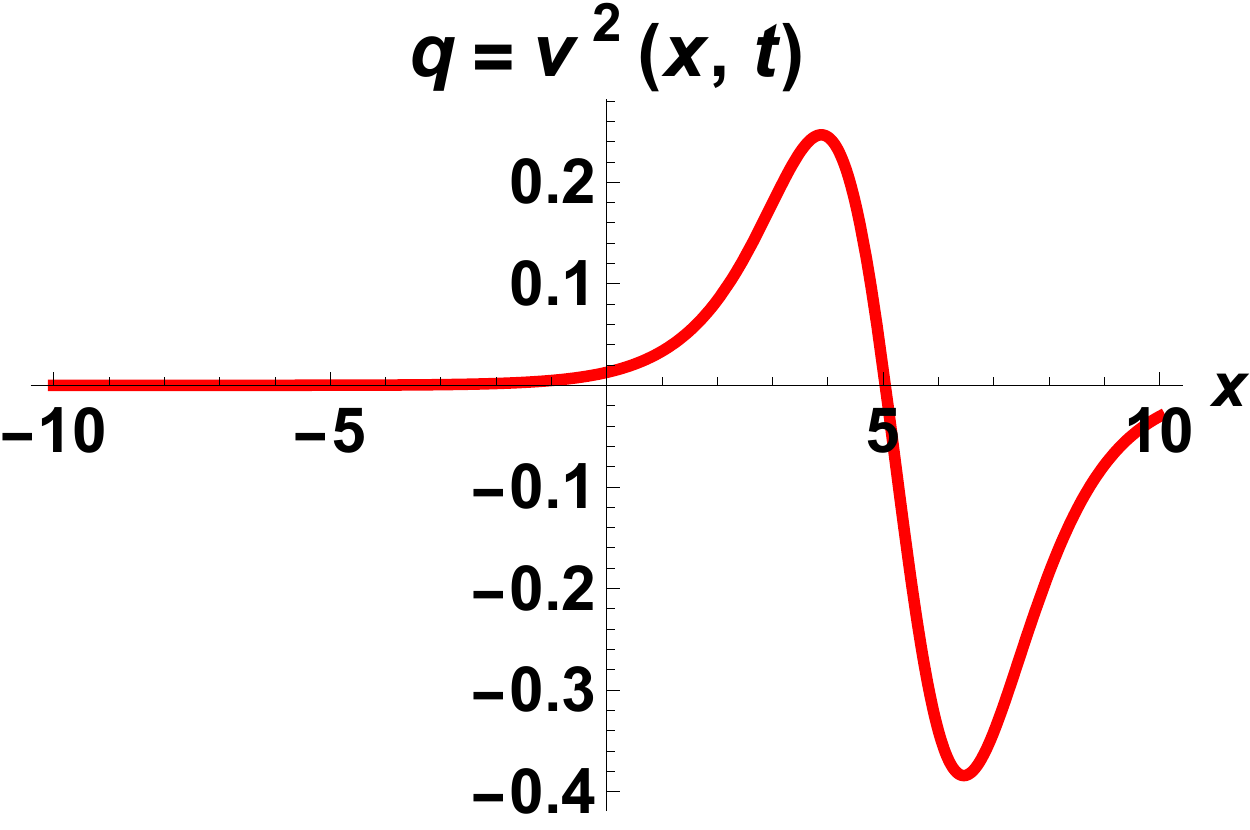}
    \caption{represents the one dimensional dynamics of two soliton solution  after the interaction}
    \label{Fig.7}
\end{figure}
Here we have calculated two-soliton solutions for the coupling field variables $u$ and $v$ which  simultaneously satisfy CNW equation (\ref{1}). The two-soliton solution $ u[2]$  can be assumed as interaction of  KdV solitons associated to CNW equation (\ref{1}), where as  $ v[2]$ which can be obtained from $ q[2]$ also two-soliton solution differs from $ u[2]$ associated to same equation (\ref{1}) that composed by soliton and anti-soliton propagating in opposite directions and interact elastically.

\subsection{Three-soliton solution  $u[3]$}
The three-fold Darboux transformation (\ref{22}) with trivial solution u=0 will take the following form 
  \begin{equation}\label{37}
      u[3]=2\frac{d^2}{dx^2}\log\left[W(\psi_1,\psi_2,\psi_3)\right],
  \end{equation}
  we can compute the value for $\psi_3$ form the linear system (\ref{2}) at $\lambda=\lambda_3 $ which is given by
  \begin{equation}\label{38}
      \psi_3(x,t)=2 \cosh{(\gamma_3)}.
  \end{equation}
  with
  \begin{equation}\label{39}
      \gamma_3=k_3 x+4 k_3 \lambda_3 t.
  \end{equation}
  Now after substituting these values in equation (\ref{37}) and simplifying, we get
   \begin{multline}\label{40}
   u[3] =[-(k_1 k_3 (-k_1^2 + k_3^2) \sinh{\gamma_1} \sinh{\gamma_2} \sinh{\gamma_3} + 
      k2 \cosh{\gamma_2} (k1 (k1^2 - k2^2) \cosh{\gamma_3}\\ \sinh{\gamma_1} + 
        k3 (k2^2 - k3^2) \cosh{\gamma_1} \sinh{\gamma_3}))^2 + (k1 (-k2^2 + k3^2) \cosh{\gamma_3} \sinh{\gamma_1} \sinh{\gamma_2}\\ 
        +
       \cosh{\gamma_1} (k2 (k1^2 - k3^2) \cosh{\gamma_2} \cosh{\gamma_3} + (-k1^2 + k2^2) k3 \sinh{\gamma_2} \sinh{\gamma_3})) \\
       (k1 (k1^2 - k2^2 -
          k3^2) (k2^2 - k3^2) \cosh{\gamma_3} \sinh{\gamma_1} \sinh{\gamma_2} + 
      \cosh{\gamma_1} (-k2 (k1^2 - k3^2)\\
      (-k1^2 + k2^2 - k3^2) \cosh{\gamma_2} \cosh{\gamma_3} - (k1^2 - k2^2) k3 (k1^2 + k2^2 - k3^2) \sinh{\gamma_2} \sinh{\gamma_3}))]\\
      /[k_1 (-k_2^2 + k_3^2) \cosh{\gamma_3} \sinh{\gamma_1} \sinh{\gamma_2} + 
   \cosh{\gamma_1}\\
   (k_2 (k_1^2 - k_3^2) 
   \cosh{\gamma_2} \cosh{\gamma_3} + k_3(-k_1^2 + k_2^2) \sinh{\gamma_2} \sinh{\gamma_3})]^2,
  \end{multline}
  which is the explicit expression of three-soliton solution for $u$ associated with CNW equation and the interactions of  these solitons have been shown below.
  \begin{figure}[H]
\begin{subfigure}{0.4\textwidth}
    \includegraphics[width=1.2\linewidth, height=6cm]{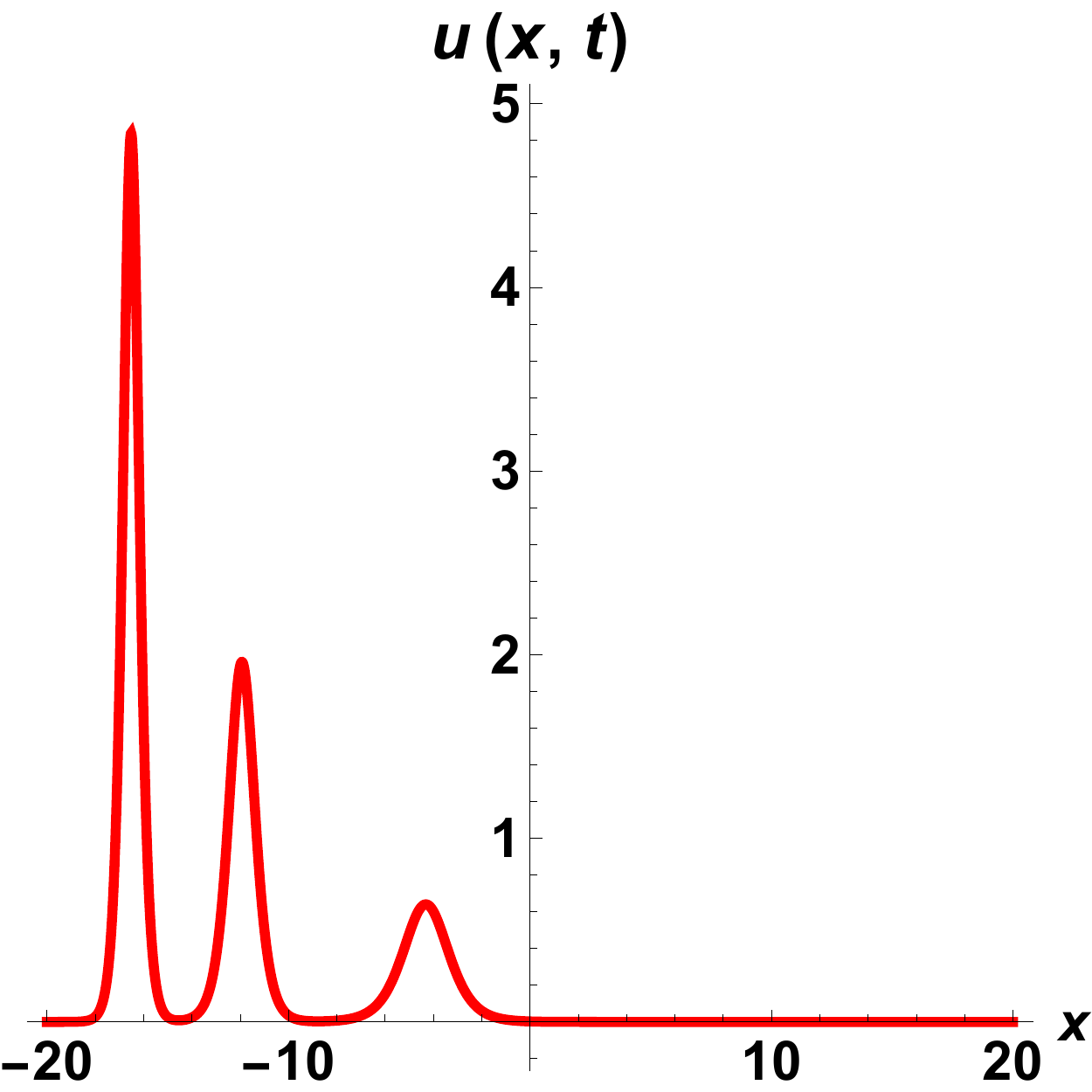}
    \caption{1-D}
    \label{Fig.8a}
\end{subfigure}
  \hspace{1cm}
    \begin{subfigure}{0.4\textwidth}
    \includegraphics[width=1.2\linewidth, height=6cm]{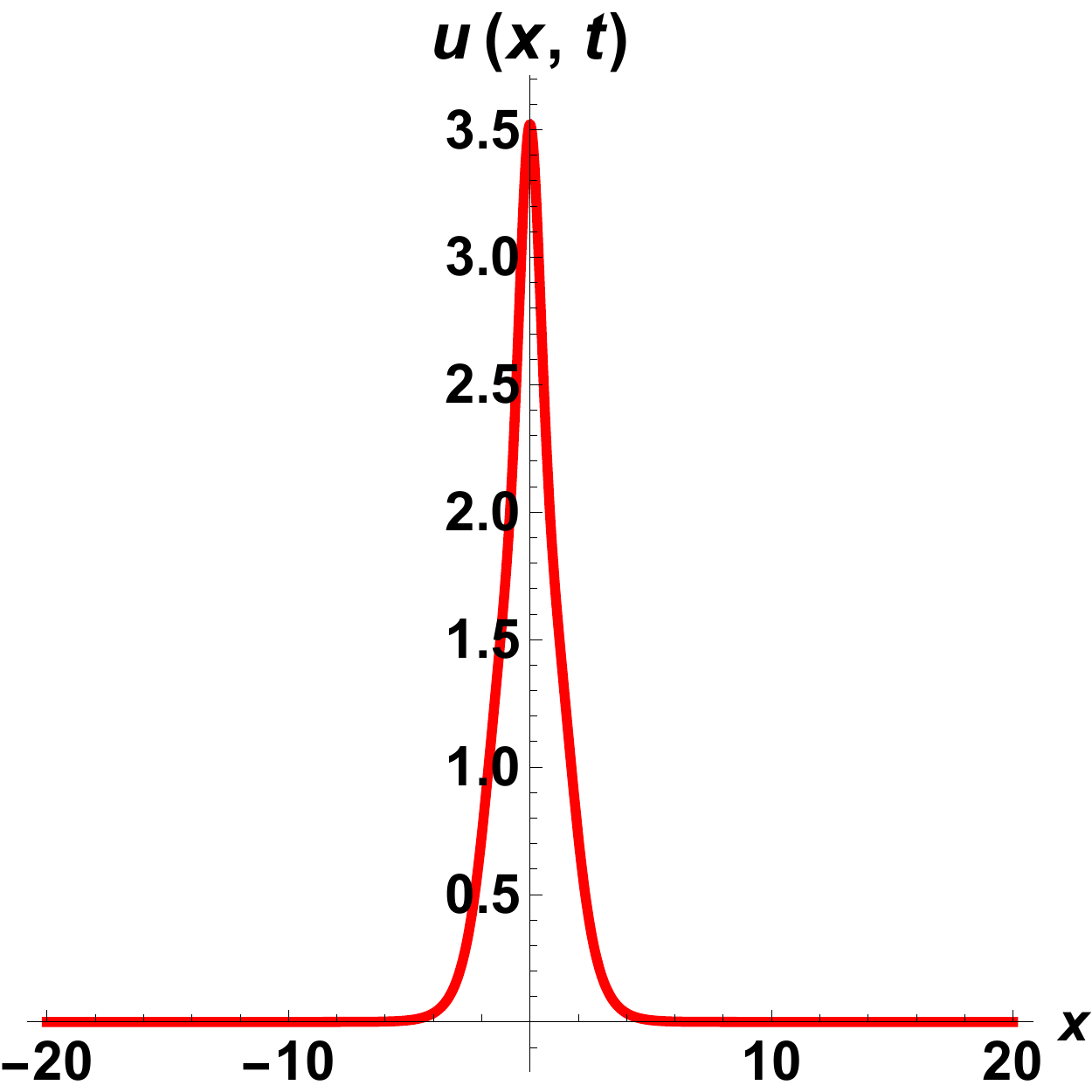}
    \caption{1-D}
    \label{Fig.8b}
    \end{subfigure}
    \caption{(a) represents the one dimensional dynamics of three soliton solution  before interaction where as in (b) one dimensional dynamics at the time of interaction has been shown.}
\label{Fig.8}
\end{figure}

\begin{figure}[H]
\begin{subfigure}{0.4\textwidth}
    \includegraphics[width=1.2\linewidth, height=6cm]{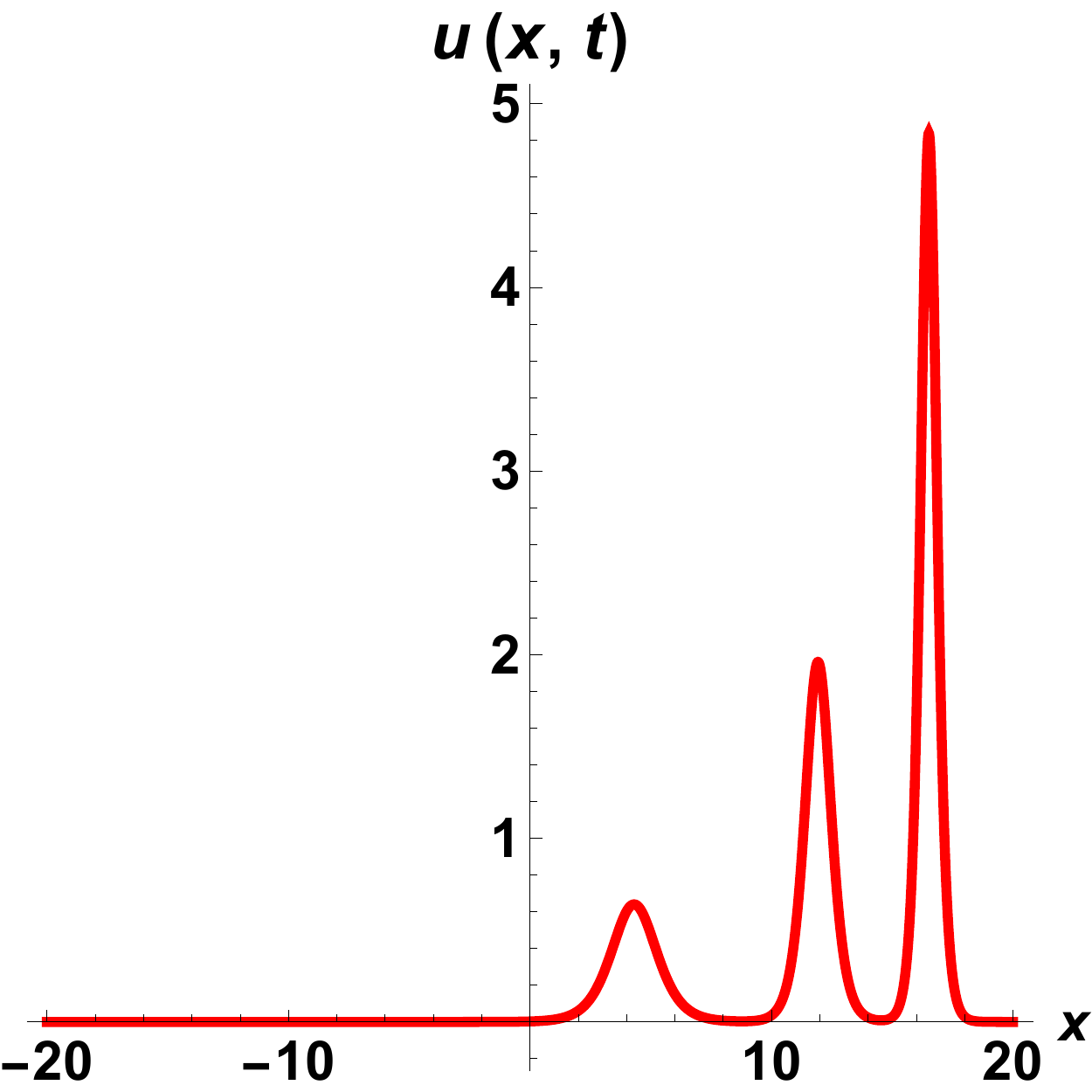}
    \caption{1-D}
    \label{Fig.9a}
\end{subfigure}
  \hspace{1cm}
    \begin{subfigure}{0.4\textwidth}
    \includegraphics[width=1.2\linewidth, height=6cm]{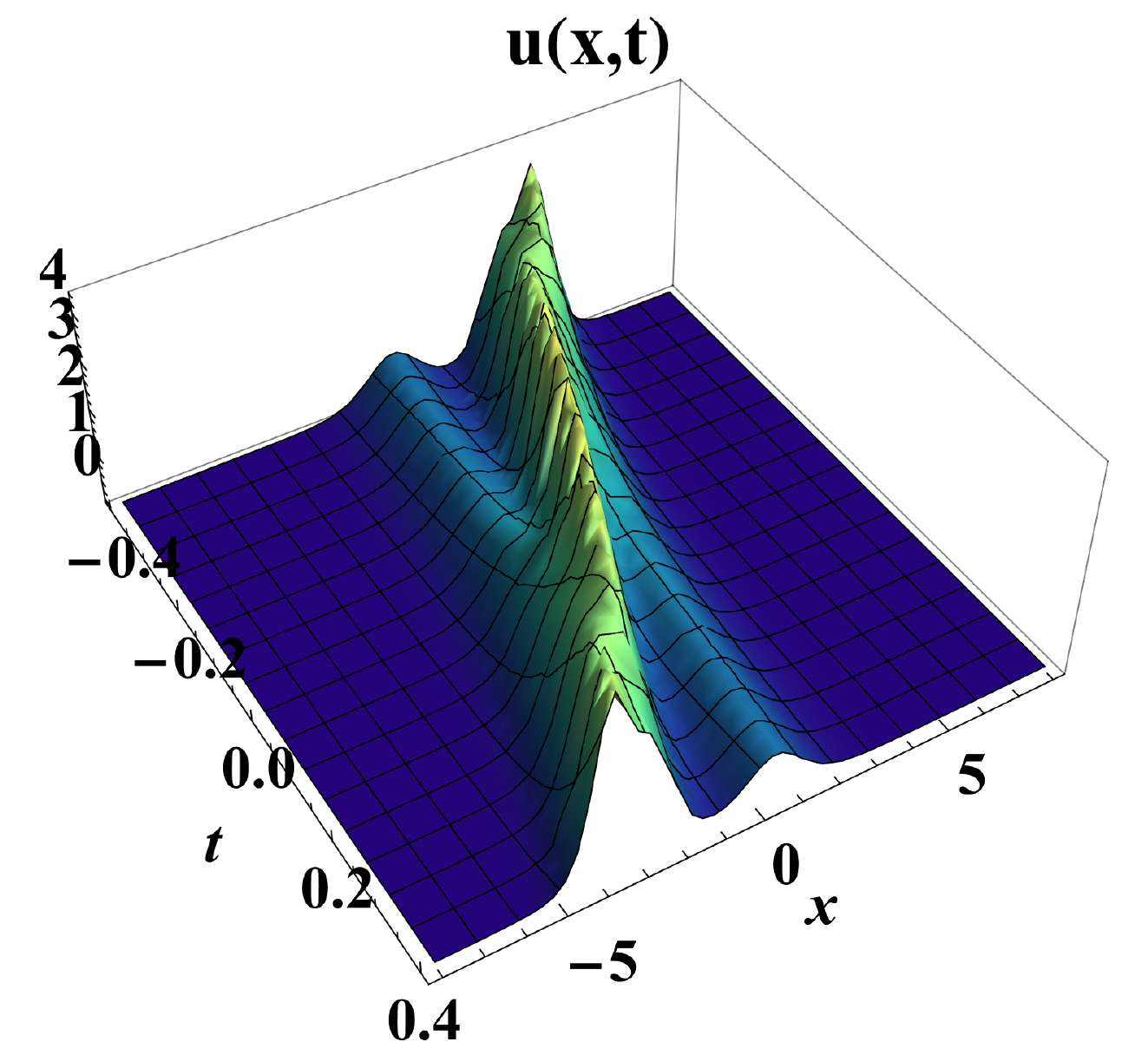}
    \caption{1-D}
    \label{Fig.9b}
    \end{subfigure}
    \caption{(a) represents the one dimensional dynamics of three soliton solution  after the interaction where as in (b) representing interaction of three solitons on plane.}
\label{Fig.9}
\end{figure}
\subsection{Three-soliton solution $q[3]$ for coupling variable $v$ }
As above we have derived KdV-type three soliton solution and it seems more substantial to calculate three soliton solution to its coupling partner $v$. For this purpose let write (\ref{23}) at $N=3$ in following form 
\begin{equation}\label{41}
    q[3]=q[2]-\left(u[2]_t-u[2]_{xxx}-6u[1]u[1]_x\right)\sigma_{3}[2]^{-1},
\end{equation}
Now substitute the calculated values of $u[2]$ and $\sigma_{3}[2]$  in above expression then after simplification, we get 
\begin{equation}\label{42}
 q[3]=q[2]-A*B
\end{equation}
where
\begin{multline}\label{43}
    A=(4 (k_1^2 - 
     k_2^2) (-\lambda_2 k_1^2 k_2^2 (k_1^2 - k_2^2) \cosh{\gamma_1} (\cosh{\gamma_1} + \cosh{\gamma_2}) \sinh{\gamma_2} \\
     (k_2 \cosh{\gamma_1} \cosh{\gamma_2} - 
        k_1 \sinh{\gamma_1} \sinh{\gamma_2})^2 + 
     4 (-k_1^2 + k_2^2 + k_2^2 \cosh{\gamma_1}+ 
        k_1^2 \cosh{\gamma_2})\\
        (k_1 k_2 (-\lambda_1 + \lambda_2) \cosh{\gamma_2} \sinh{\gamma_1} + (k_1^2 \lambda_1 - k_2^2 \lambda_2) \cosh{\gamma_1} \sinh{\gamma_2})\\
        (k_2 \cosh{\gamma_1} \cosh{\gamma_2} - 
        k_1 \sinh{\gamma_1} \sinh{\gamma_2})^2 - 
     6 k_1 k_2 (k_1^2 - 
        k_2^2) \\
        (k_2 \cosh{\gamma_1}\cosh{\gamma_2}- 
        k_1 \sinh{\gamma_1} \sinh{\gamma_2}) (1/
         2 \cosh{\gamma_1}^2 (-3 k_1^2 +\\ 
           4 k_2^2 + (3 k_1^2 - 2 k_2^2) \cosh{2 \gamma_2}) - 
        k_1^2 \sinh{\gamma_1}^2 \sinh{\gamma_2}^2) (k_2 \sinh{2 \gamma_1}+ 
        k_1 \sinh{2 \gamma_2})\\
        + 
     4 k_1^2 k_2^2 (-k_2 \cosh{\gamma_1}\cosh{\gamma_2} + 
        k_1 \sinh{\gamma_1} \sinh{\gamma_2})^3 (k_1 \sinh{2 \gamma_1} + k_2 \sinh{2 \gamma_2} +\\ 
     4 k_1 k_2 (k_2 \cosh{\gamma_1} \cosh{\gamma_2} - 
        k_1 \sinh{\gamma_1} \sinh{\gamma_2})^3 (k_2 \lambda_1 \sinh{2 \gamma_1} + 
        k_1 \lambda_2 \sinh{2 \gamma_2}) +\\
        (k_1^2 - k_2^2) (k_1^2 - k_2^2 - 
        k_2^2 \cosh{2 \gamma_1} - 
        k_1^2 \cosh{2 \gamma_2}) (-k_1 k_2 \cosh{\gamma_1}^2 \cosh{\gamma_2} (-k_1^2 - 2 k_2^2 +\\ 
           k_1^2 \cosh{2 \gamma_2}) \sinh{\gamma_1} +
        \cosh{\gamma_1}^3 (-6 k_1^4 + 17 k_1^2 k_2^2 - 
           10 k_2^4 + (6 k_1^4 - 7 k_1^2 k_2^2 + 2 k_2^4) \cosh{2 \gamma_2})\\ \sinh{\gamma_2} - 
        k_1^2 \cosh{\gamma_1} (-4 k_1^2 + 
           7 k_2^2 + (4 k_1^2 - 3 k_2^2) \cosh{2 \gamma_2}) \sinh{\gamma_1}^2 \sinh{\gamma_2} +\\
        k_1^3 k_2 \sinh{\gamma_1}^3 \sinh{\gamma_2} \sinh{2 \gamma_2}) + 
     6 (k_1^2 - k_2^2) (k_1^2 - k_2^2 - k_2^2 \cosh{2 \gamma_1} -
        k_1^2 \cosh{2 \gamma_2})\\ (-2 (k_1^2 - k_2^2) \cosh{\gamma_1} (-k_1^2 + k_2^2 + 
          k_2^2 \cosh{2 \gamma_1} +
           k_1^2 \cosh{2\gamma_2}) \sinh{\gamma_2} + 
        2 k_1 k_2 \\
        (-k_2 \cosh{\gamma_1} \cosh{\gamma_2} + 
           k_1 \sinh{\gamma_1}\sinh{\gamma_2}) (k_2 \sinh{2 \gamma_1} + 
           k_1 \sinh{2 \gamma_2}))))\\
           /(k_2 \cosh{\gamma_1} \cosh{\gamma_2} - 
   k_1 \sinh{\gamma_1} \sinh{\gamma_2})^5,
\end{multline}
\begin{multline}\label{44}
    B=(k_1 (-k_2^2 + k_3^2) \cosh{\gamma_3} \sinh{\gamma_1} \sinh{\gamma_2} + 
   \cosh{\gamma_1} (k_2 (k_1^2 - k_3^2) \cosh{\gamma_2} \cosh{\gamma_3}\\
   + (-k_1^2 + k_2^2) k_3 \sinh{\gamma_2} \sinh{\gamma_3}))/(k_1 k_3 (-k_1^2 + 
      k_3^2) \sinh{\gamma_1} \sinh{\gamma_2} \sinh{\gamma_3} +\\ 
   k_2 \cosh{\gamma_2} (k_1 (k_1^2 - k_2^2) \cosh{\gamma_3} \sinh{\gamma_1} + 
      k_3 (k_2^2 - k_3^2) \cosh{\gamma_1}\sinh{\gamma_3})),
\end{multline}
this is straight forward to obtain $ v[3]$ by taking square root of $ q[3]$ and we have shown  three soliton elastic interaction in one dimension.
\begin{figure}[H]
\begin{subfigure}{0.4\textwidth}
    \includegraphics[width=5cm, height=5.5cm]{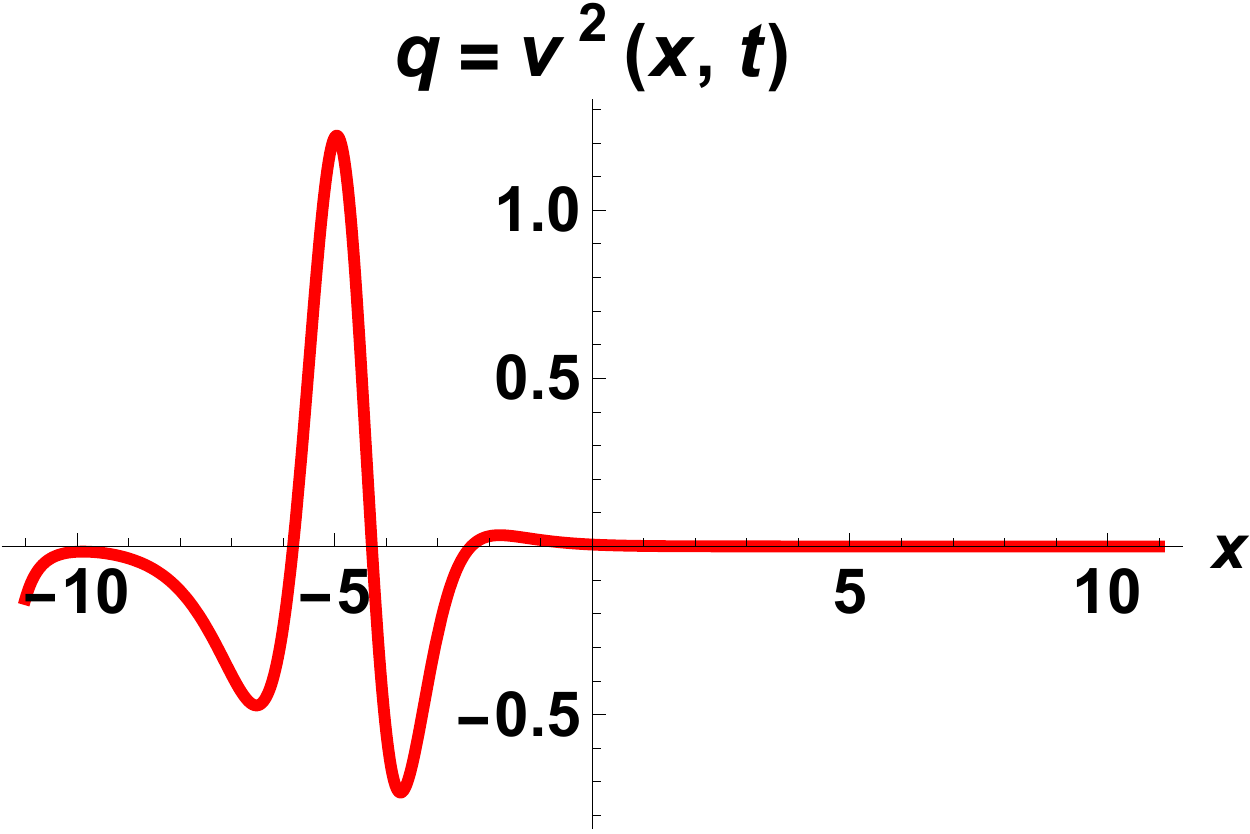}
    \caption{1-D}
    \label{Fig.10a}
\end{subfigure}
  \hspace{1cm}
    \begin{subfigure}{0.4\textwidth}
    \includegraphics[width=1.2\linewidth, height=6cm]{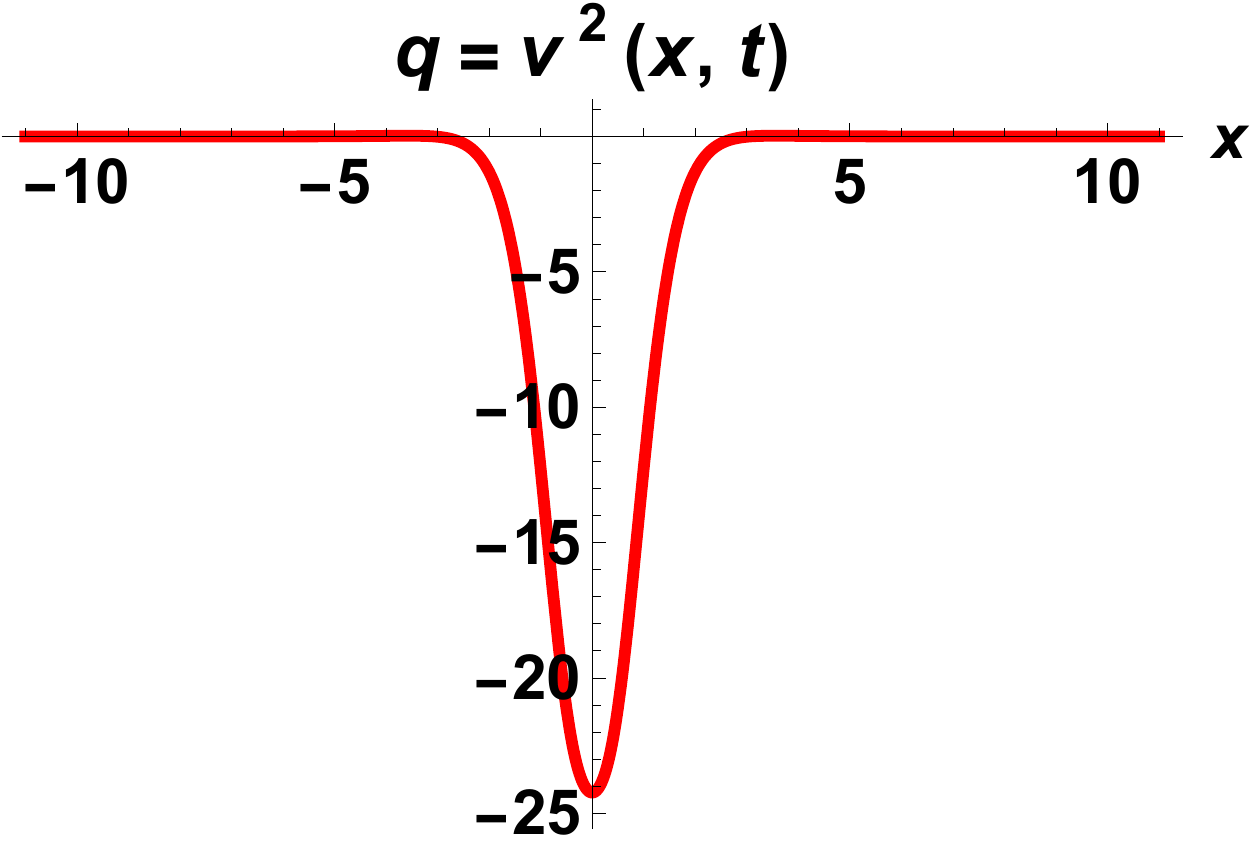}
    \caption{1-D}
    \label{Fig.10b}
    \end{subfigure}
    \caption{(a) represents the one dimensional dynamics of three soliton solution  before interaction where as in (b) one dimensional dynamics at the time of interaction has been shown.}
\label{Fig.10}
\end{figure}

\begin{figure}[H]
    \centering
    \includegraphics[width=6cm, height=6cm]{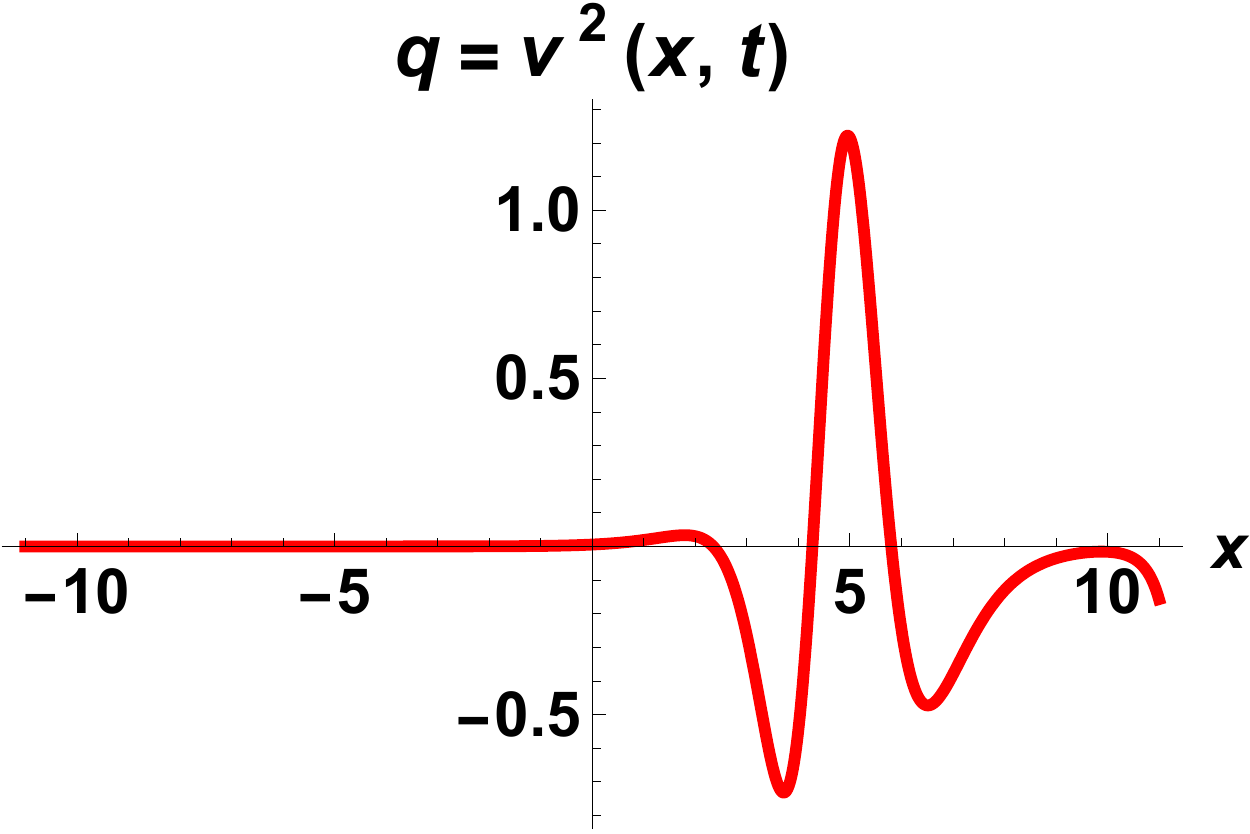}
    \caption{represents the one dimensional dynamics of three soliton solution  after the interaction}
    \label{Fig.11}
\end{figure}

As in above last section, we derived three-soliton solutions for the coupling field variables $u$ and $v$ which  simultaneously satisfy CNW equation (\ref{1}). The three-soliton solution $ u[3]$  can be assumed as interaction of  KdV solitons associated to CNW equation (\ref{1}), where as  $ v[3]$ which can be obtained from $ q[3]$ also three-soliton solution differs from $ u[3]$ associated to same equation (\ref{1}) that composed by soliton and anti-soliton propagating in opposite directions and interact elastically.

\section{Equation of Continuity and conserved densities}
This section is devoted to construct the equation of continuity which is associated with CNW equation (\ref{1}) and derivation of  conserved densities incorporating its riccati equation.\\
For this purpose, let define a quantity $\Gamma$ in terms of arbitrary function, as below
\begin{equation}\label{45}
    \Gamma=\frac{\psi^{\prime}}{\psi}=(ln\psi)_x,
\end{equation}
Now take the time derivative of $\Gamma$  and then make use of equation (\ref{3}) in resulting expression, we get folowing result after simplification 
 \begin{equation}\label{46}
    \frac{\partial \Gamma}{\partial t}=2 u^{\prime}\Gamma+\left(4 \lambda+2u\right)\left(\frac{\psi^{\prime \prime}}{\psi}-\frac{{\psi^{\prime}}^2}{\psi^2}\right)-u^{\prime \prime},
\end{equation}
and the space derivative of equation(\ref{45}) can be written as 
\begin{equation}\label{47}
    \frac{\partial \Gamma}{\partial x}=\frac{\psi^{\prime \prime}}{\psi}-\frac{{\psi^\prime}^2}{\psi^2},
\end{equation}
Now eliminating $\left(\frac{\psi^{\prime \prime}}{\psi}-\frac{{\psi^{\prime}}^2}{\psi^2}\right)$ from equation (\ref{46}), and by using last equation  equation(\ref{47}), after simplification we obtain resulting expression as below 
\begin{equation}\label{48}
    \frac{\partial \Gamma}{\partial t}=\frac{\partial}{\partial x}\left[(4 \lambda+2u)\Gamma-u^{\prime}\right],
\end{equation}
or
\begin{equation}\label{49}
    \frac{\partial \rho}{\partial t}-\frac{\partial J}{\partial x}=0
\end{equation}
the above is the equation of continuity associated with CNW equation (\ref{1}) involving  density as $\rho$ and corresponding  current as $j$ given as 
\begin{equation}\label{50}
    J=(4 \lambda+2u)\Gamma-u^{\prime}.
\end{equation}
We may construct the hierarchy of conserved densities for CNW equation (\ref{1}) with the help of its riccati equation as calculated below.\\
It is straight forward to construct the riccati  equation from expression  (\ref{47}) by using equation (\ref{2}) as in the following form

\begin{equation}\label{51}
    \frac{\partial\Gamma}{\partial x}+\left(u-\lambda+\frac{q}{4 \lambda}\right)+\Gamma^2=0,
\end{equation}
Now by taking the derivation of above Riccati equation with respect $x$, we obtain second order ordinary differential equation as 
\begin{equation}\label{52}
    \frac{\partial^2 \Gamma}{\partial x^2}-2\Gamma^3+2 \Gamma \left(\lambda-u-\frac{q}{4\lambda}\right)+u^{\prime}+\frac{q^{\prime}}{4\lambda}=0.
\end{equation}
Let consider following ansatz as the solution of above equation 
\begin{equation}\label{53}
    \Gamma=\sum_{n=1}^{\infty}\Gamma_n \lambda^{-n},
\end{equation}
where the quantities $\Gamma_n $ are to be determined which represent the densities associated with Ito type coupled equation (\ref{1}).   After substituting ansatz (\ref{53}) into equation(\ref{52}) and then expanding summation, finally the coefficients of  $\lambda^{0} $,  $\lambda^{-1} $ , $\lambda^{-2} $   in resulting expression produce first three densities 
\begin{equation}\label{54}
    \Gamma_1=-\frac{u^{\prime}}{2},
\end{equation}
\begin{equation}\label{55}
    \Gamma_2=\frac{u^{\prime \prime \prime}}{4}-\frac{q^{\prime}}{8}-\frac{1}{2}uu^{\prime},
\end{equation}
\begin{equation}\label{56}
    \Gamma_3=u \Gamma_2+\frac{q}{4}\Gamma_1-\frac{1}{2}\frac{\partial^2}{\partial x^2} \Gamma_2.
\end{equation}
 and the coefficients  of $N$-the terms yield the following relation
\begin{equation}
    \Gamma_{n+1}=\frac{1}{2}\left[2u\Gamma_n+\frac{q}{2}\Gamma_{n-1}+2 \sum_{l=1}^{m=n-2} \Gamma_l\left(\sum_{k=1}^{j=m-l+1}\Gamma_k \Gamma_{j-k+1}\right)-\frac{\partial^2}{\partial x^2} \Gamma_n \right],
\end{equation}
which holds for $n=3,4,5,..., $ and with the help of that expression  we may calculate all  remaining conserved densities. Now we can  expression  (\ref{47})  can be written in $n$-th form $  J_n=(4 \lambda+2u)\Gamma_n-u^{\prime}$that yields the corresponding currents.
\section{Conclusion}
    In this article, we have calculated multi-solitonic solutions for  coupling field variables $u$, $q$ associted with  Ito type coupled KdV equation (\ref{1}) in Darboux framework and then   their $N$-fold Darboux  transformations  generalised in terms of Wronkians. We also derived its equation of continuity with hierarchy of conserved densities. For  motivation, It is quite interesting to construct its Hirota bilinear form with multi-solitoinc solutions and comparable to our results obtained here, that will be discussed in a separate   paper.  Further, it is straight forward to calculate its noncommutative  analogue with  $N$-fold Darboux solutions in terms of quasideterminats which may coincide with our results under commutative limit. 

\section{Acknowledgement}
This research work  has been completed as the part of Belt and  Road Young Scientist project sponsored by  Science and technology commission of Shanghai at college of science, Shanghai University with project, No. $20590742900$. We are also thankful  to the Punjab University 54590 on providing me the facilities to complete that work.

\end{document}